\documentstyle[aps,prl,epsf]{revtex}

\newcommand{\pdag}{{\phantom{\dagger}}}
\newcommand{\bq}{\begin{equation}}
\newcommand{\eq}{\end{equation}}
\newcommand{\bn}{\begin{eqnarray}}
\newcommand{\en}{\end{eqnarray}}

\begin{document}
\draft
\wideabs{
\title{Nonequilibrium Kondo Effect in a Multi-level Quantum Dot near singlet-triplet 
transition}
\author{Bing Dong and X. L. Lei}
\address{Department of Physics, Shanghai Jiaotong University, 1954 Huashan Road, Shanghai 
200030, China}
\maketitle
\begin{abstract}
The linear and nonlinear transport through a multi-level lateral quantum dot connected to 
two leads is investigated using a generalized finite-$U$ slave-boson mean field approach. 
For a two-level quantum dot, our calculation demonstrates a substantial conductance 
enhancement near the degeneracy point of the spin singlet and triplet states, a 
non-monotonic temperature-dependence of conductance and a sharp dip and nonzero bias 
maximum of the differential conductance. These agree well with recent experiment 
observations. This two-stage Kondo effect in an out-of-equilibrium situation is attributed 
to the interference between the two energy levels.  

\end{abstract}
 }

Since the observation of the Kondo effect in semiconductor quantum dot (QD) with odd 
electron number \cite{Goldhaber}, there have been a great deal of experimental and 
theoretical investigation into this many-body phenomenon. With easy control of major 
parameters of these artificial atoms in a wide range, QDs facilitate exploration of the 
Kondo physics at very different environments. As a consequence, new types of Kondo effects 
have been discovered in experiments. Very recently, a surprising Kondo-enhanced 
conductance has also been detected for an even number of electrons in both vertical 
\cite{Sasaki} and lateral \cite{Schmid,Wiel} QD configurations, which is markedly 
different from the conventional spin-$1/2$ Kondo effect.   

A real QD contains more than one energy levels and usually two electrons incline to occupy 
the lower level, resulting in a singlet local spin state. Decreasing the spacing of the 
two energy levels by applying a magnetic field makes it possible for two electrons to 
occupy two different levels due to the exchange interaction $J$, forming the local 
spin-triplet state. Scaling theory \cite{vertical,lateral} and numerical renormalization 
group analysis \cite{Izumida,Hofstetter} revealed that this singlet-triplet transition 
gives rise to an anomalous enhancement of conductance at low temperature. 

The physics in the lateral and vertical QD configurations are different. Recent 
observation on the lateral QD has disclosed a two-stage Kondo effect in the 
temperature-dependent conductance and differential conductance, and favored an 
interpretation in terms of a single conduction channel in two leads. \cite{Wiel} So far, 
there have been only a scaling analysis \cite{lateral} and a numerical renormalization 
group calculation \cite{Hofstetter} on this problem in the low bias regime. Its clear 
physical picture and the nonlinear conductance at finite bias need further exploring. In 
this letter we investigate the singlet-triplet Kondo effect in a lateral QD having two 
energy levels in an out-of-equilibrium situation, by generalizing the finite-$U$ 
slave-boson mean-field (SBMF) approach, which was initially developed by 
Kotliar-Ruckenstein \cite{KR} and extended by us to study Kondo-type transport through QD 
\cite{Dong1} and coupled QDs \cite{Dong2}. Our results manifest not only a substantial 
enhancement of conductance near the local spin singlet-triplet transition, but also a 
two-stage Kondo effect in temperature-dependent conductance and differential conductance.     

The Hamiltonian of the QD with two orbital energy levels ($j=1\,,2$) connected to two 
leads ($\eta=L/R$) can be written as $H=H_{L}+H_{R}+H_{D}+H_{T}$, in which 
$H_{\eta=}\sum_{\eta,k,\sigma}\epsilon _{\eta k }
c_{\eta k \sigma }^{\dagger }c_{\eta k \sigma }^\pdag$ and $H_{T}=\sum_{j,\eta, k, 
\sigma}(V_{j\eta}c_{\eta k \sigma }^{\dagger }c_{j\sigma}^\pdag+{\rm {H.c.}})$ describe 
the left and right leads and the tunneling between each lead and the QD (assuming a single 
conduction channel available per lead), respectively, and $H_{D}$ represents the isolated 
QD ($\alpha,\beta,\gamma,\sigma=\pm 1$):
\begin{eqnarray}
H_{D}&=&\sum_{j,\sigma
}\epsilon_{j}c_{j\sigma }^{\dagger }c_{j\sigma }^\pdag+\sum_{j}U_j n_{j\uparrow 
}n_{j\downarrow }+U_{12}\sum_{\alpha\beta}n_{1\alpha}n_{2\beta} \cr
&\hphantom{=}& -J\sum_{\alpha\beta\gamma\sigma} c_{1\alpha}^{\dagger}c_{1\beta}^\pdag 
c_{2\gamma}^{\dagger}c_{2\sigma}^\pdag{\bbox \sigma}_{\alpha\beta}\cdot {\bbox 
\sigma}_{\gamma\sigma},
\label{dothamiltonian}
\end{eqnarray}
where $\epsilon_{j}$ in the first term is the single particle energy, the second and third 
terms denote the intra- and inter-level Coulomb interactions, and the last term gives a 
ferromagnetic exchange coupling $J>0$ due to Hund's rule. The energy spacing 
$\delta=\epsilon_2-\epsilon_1$ can be controlled by an external magnetic field. However, 
considering the small $g$-factor in, for example, GaAs, we neglect the Zeeman splitting in 
the QD, as done in previous theoretical treatment. \cite{Izumida,Hofstetter} To keep the 
discussion simple we assume $U_{j}=U_{12}=U$ and $V_{j\eta}=V$ ($j=1\,,2$ and 
$\eta=L\,,R$).      

Following the SBMF scheme, \cite{KR} we introduce sixteen auxiliary Bose fields associated 
with sixteen single-particle eigenstates of the isolated multi-level QD Hamiltonian 
(\ref{dothamiltonian}), as shown in Table \ref{t1}. For example, boson operator 
$d_{1S_{z}}$ is with the spin-triplet state having spin $(1,S_{z})$, and $d_{01}$ is with 
the spin-singlet state $|\uparrow \downarrow,0\rangle$ (it is the lowest energy state in 
all three spin-singlet states). Note that the splitting of energies between this 
spin-singlet state and spin-triplet state $\Delta=J-\delta$ can be tuned by a magnetic 
field. Therefore, for $\delta=J$ the three triplet states $d_{1S_{z}}$ and the singlet 
state $d_{01}$ are degenerate. Necessarily, the completeness relation for these 
slave-boson operators, $I=1$ ($I\equiv e^\dagger e+\sum_{j\sigma}p_{j\sigma}^\dagger 
p_{j\sigma}^{\pdag}+\sum_{S_{z}\in \{1,0,\bar{1}\}}d_{1S_{z}}^\dagger 
d_{1S_{z}}^{\pdag}+\sum_{l=1}^{3}d_{0l}^\dagger d_{0l}^\pdag
+\sum_{j\sigma}t_{j\sigma}^\dagger t_{j\sigma}^\pdag+f^\dagger f$), and the condition for 
the correspondence between fermions and bosons, $n_{j\sigma}\equiv c_{j\sigma}^\dagger 
c_{j\sigma}^\pdag=Q_{j\sigma}$ ($Q_{j\sigma}\equiv p_{j\sigma}^\dagger 
p_{j\sigma}^\pdag+d_{1\sigma}^\dagger d_{1\sigma}^\pdag+\frac{1}{2} d_{10}^\dagger 
d_{10}^\pdag + d_{0j}^\dagger d_{0j}^\pdag
+ \frac{1}{2}d_{03}^\dagger d_{03}^\pdag + t_{j\sigma}^\dagger t_{j\sigma}^\pdag + 
\sum_{\sigma '}t_{\bar{j}\sigma'}^\dagger t_{\bar{j} \sigma'}^\pdag +f^\dagger f$ 
[$\sigma,\sigma'=\pm 1$]), should be imposed to confine the enlarged Hilbert space. 
Moreover, in the combined fermion-boson representation, the QD fermion operators 
$c_{j\sigma}^\dagger$ and $c_{j\sigma}$ in the hopping term are exppressed as
$z_{j\sigma}^\dagger c_{j\sigma}^\dagger$ and
$c_{j\sigma} z_{j\sigma}$, respectively, where $z_{j\sigma}$ consists of all the boson 
operator sets which are associated with the physical process that a $\sigma$-spin electron 
of the $j$th level is annihilated ($\bar{j}\neq j$, $\bar{\sigma}\neq\sigma$):
\begin{eqnarray}
z_{j\sigma}&=&Q_{j\sigma}^{-1/2}\left (1-Q_{j\sigma} \right )^{-1/2}\left[ e^\dagger 
p_{j\sigma}^\pdag + p_{\bar{j}\sigma}^\dagger d_{1\sigma}^\pdag \right .\cr 
&\hphantom{=}&+ \frac{1}{2} p_{\bar{j}\bar{\sigma}}^\dagger \left ( d_{10}^\pdag + 
d_{03}^\pdag \right ) + p_{j\bar{\sigma}}^\dagger d_{0j}^\pdag + d_{0\bar{j}}^\dagger 
t_{j\sigma}^\pdag \cr
&\hphantom{=}&\left. + \frac{1}{2} \left(d_{10}^\dagger + d_{03}^\dagger \right ) 
t_{\bar{j}\sigma}^\pdag + d_{1\bar{\sigma}}^\dagger t_{\bar{j}\bar{\sigma}}^\pdag + 
t_{j\bar{\sigma}}^\dagger f\right ].
\end{eqnarray}

\begin{figure}
\vspace*{8.3cm}
\smash{\vbox{\begin{table}
\caption{
Sixteen Eigenstates, spin quantum numbers $S$, $S_{z}$ and energies $E$ for the isolated 
QD with two levels $1$ and $2$, and the assigned slave-boson (SB) operators.}
\label{t1}
\begin{tabular}{cccc} 
Eigenstate & ($S$, $S_{z}$) & $E$ & SB \\ \hline
$|0,0\rangle $ & ($0$,  $0$) & $0$ & $e$ \\ 
$|\uparrow ,0\rangle $ & ($1/2$,  $1/2$) & $\epsilon_1$ & $p_{1\uparrow }$ \\ 
$|\downarrow ,0\rangle $ & ($1/2$, $-1/2$) & $\epsilon_1$ & $p_{1\downarrow }$ \\ 
$|0,\uparrow \rangle $ & ($1/2$, $1/2$) & $\epsilon_2$ & $p_{2\uparrow }$ \\ 
$|0,\downarrow \rangle $ & ($1/2$, $-1/2$) & $\epsilon_2$ & $p_{2\downarrow }$ \\ 
$|\uparrow ,\uparrow \rangle $ & ($1$,  $1$) & $\epsilon_1+\epsilon_2+U-J$ & $d_{11}$ \\ 
$\frac{1}{\sqrt{2}}\left( |\uparrow ,\downarrow \rangle +|\downarrow ,\uparrow \rangle 
\right)$ & ($1$,  $0$) & $\epsilon_1+\epsilon_2+U-J$ & $d_{10}$ \\ 
$|\downarrow ,\downarrow \rangle $ & ($1$,  $-1$) & $\epsilon_1+\epsilon_2+U-J$ & 
$d_{1\bar{1}}$ \\ 
$|\uparrow \downarrow ,0\rangle$ & ($0$, $0$) & $2\epsilon_1+U$ & $d_{01}$ \\ 
$|0,\uparrow \downarrow \rangle$ & ($0$, $0$) & $2\epsilon_2+U$ & $d_{02}$ \\ 
$\frac{1}{\sqrt{2}}\left( |\uparrow ,\downarrow \rangle -|\downarrow ,\uparrow \rangle 
\right)$ & ($0$, $0$) & $\epsilon_1+\epsilon_2+U+3J$ & $d_{03}$ \\ 
$|\uparrow ,\uparrow \downarrow \rangle $ & ($1/2$, $1/2$) & $2\epsilon_1+\epsilon_2+3U-J$ 
& $t_{1\uparrow}$ \\ 
$|\downarrow ,\uparrow \downarrow \rangle $ & ($1/2$, $-1/2$) & 
$2\epsilon_1+\epsilon_2+3U-J$ & $ t_{1\downarrow }$ \\ 
$|\uparrow \downarrow ,\uparrow \rangle $ & ($1/2$, $1/2$) & $\epsilon_1+2\epsilon_2+3U-J$ 
& $t_{2\uparrow}$ \\ 
$|\uparrow \downarrow ,\downarrow \rangle $ & ($1/2$, $-1/2$) & 
$\epsilon_1+2\epsilon_2+3U-J$ & $t_{2\downarrow }$ \\ 
$|\uparrow \downarrow ,\uparrow \downarrow \rangle $ & ($0$, $0$) & 
$2\epsilon_1+2\epsilon_2+4U-2J$ & $f$ \\ 
\end{tabular}
\end{table}}}
\vspace{-8mm}
\end{figure}

\noindent Within the slave-boson scheme, the Hamiltonian of the system can be replaced by 
the following effective Hamiltonian in terms of auxiliary boson operators plus the 
constraints incorporated via the Lagrange multipliers $\lambda$ and $\lambda_{j\sigma}$:
\begin{eqnarray}
H_{\rm eff}&=&\sum_{\eta, k,\sigma}\epsilon _{\eta k}
c_{\eta k\sigma }^{\dagger }c_{\eta k\sigma }^\pdag + \sum_{j,\sigma} 
\epsilon_{j}c_{j\sigma }^{\dagger }c_{j\sigma }^\pdag \cr
&\hphantom{=}&\hspace{-0.5cm} + U\sum_{j} d_{0j}^\dagger d_{0j}^\pdag + 
(U-J)\sum_{S_{z}\in \{ {1,0,-1}\}}d_{1S_{z}}^\dagger d_{1S_{z}}^\pdag  \cr
&\hphantom{=}&\hspace{-0.5cm} + (U+3J)d_{03}^\dagger d_{03}^\pdag + 
(3U-J)\sum_{j,\sigma}t_{j\sigma}^\dagger t_{j\sigma}^\pdag \cr
&\hphantom{=}&\hspace{-0.5cm} + (4U-2J)f^\dagger f + V\sum_{j, \eta, k, \sigma} \left (
c_{\eta k \sigma }^{\dagger } c_{j\sigma }^\pdag z_{j\sigma}^\pdag + {\rm {H.c.}} \right) 
\cr 
&\hphantom{=}&\hspace{-0.5cm} + \lambda (I-1 )+\sum_{j,\sigma }\lambda_{j\sigma } 
(c_{j\sigma }^{\dagger} c_{j\sigma }^\pdag - Q_{j\sigma}^\pdag).
\label{hamiltonian}
\end{eqnarray}

From the effective Hamiltonian (\ref{hamiltonian}) one can derive equations of motion
of the slave-boson operators. Then we use the mean-field approximation in these equations 
and in the constraints, in which all the boson operators are replaced by their expectation 
values. The negligible Zeeman splitting helps to reduce the number of the independent 
expectation values, such that we have $p_{j\sigma}=p_{j}$, $d_{1S_{z}}=d_{1}$, 
$t_{j\sigma}=t_{j}$, $z_{j\sigma}=z_{j}$, and $\lambda_{j\sigma}=\lambda_{j}$. With the 
help of the Langreth analytical continuation rules for the close time-path Green's 
function (GF), \cite{Langreth} these equations of motion can be closed in terms of the 
distribution GF $G_{j}^{<}(\omega)$ of the QD, leading to the following self-consistent 
set of equations ($j,j'=1,2$):   
\begin{eqnarray}
&&e^{2} + 2\sum_{j} p_{j}^{2} + 3d_{1}^2 + \sum_{l=1}^{3}d_{0l}^{2}+ 
2\sum_{j}t_{j}^{2}+f^{2}=1, \label{set1c} \\
&&\frac{1}{2\pi i}\int d\omega G_{j}^{<}(\omega )=p_{j}^{2} + \frac{1}{2}( d_{03}^2 + 
3d_{1}^{2}) + d_{0j}^{2}\cr 
&&\hspace{4.8cm}+t_{j}^{2}+2t_{\bar{j}}^{2}+f^{2},  \label{set2c} \\
&&\sum_{j }\frac{\partial \ln z_{j }}{
\partial e} {\cal P}_{j}+\lambda e =0, \\
&& \sum_{j'} \frac{\partial \ln z_{j' }}{\partial p_{j}}
{\cal P}_{j'} + 2\left( \lambda -\lambda _{j}\right) p_{j }=0, \\
&& \sum_{j} \frac{\partial \ln z_{j}}{\partial d_{1}} {\cal P}_{j} + 3\left( \lambda 
-\lambda _{1}-\lambda_{2} + U-J\right) d_{1} =0,  \label{set3c} \\
&& \sum_{j'} \frac{\partial \ln z_{j' }}{\partial d_{0j}} {\cal P}_{j'} + 
(\lambda-2\lambda_{j}+U)d_{0j}=0, \label{set4c} \\
&& \sum_{j} \frac{\partial \ln z_{j }}{\partial d_{03}} {\cal P}_{j} + 
(\lambda-\lambda_1-\lambda_2+U+3J)d_{03}=0, \label{set5c} \\
&& \sum_{j'} \frac{\partial \ln z_{j' }}{\partial t_{j}} {\cal P}_{j'} + 
2(\lambda-\lambda_{j}-2\lambda_{\bar{j}}+3U-J)t_{j}=0, \label{set6c} \\
&& \sum_{j} \frac{\partial \ln z_{j }}{\partial f} {\cal P}_{j} + 
2(\lambda-2\lambda_1-2\lambda_2+4U-2J)f=0, \label{set7c}
\end{eqnarray}
in which ${\cal P}_{j}=\frac{1}{2\pi i}\int d\omega G_{j }^{<}(\omega )\left( \omega 
-\tilde{\epsilon}_{j}\right)$ and
\bq
G_{j}^<(\omega)=i \frac { \tilde{\Gamma}_{j} [f_{L}(\omega)+f_{R}(\omega)]
(\omega-\tilde{\epsilon}_{\bar{j}})^2}{|{\cal D}(\omega)|^2},
\eq
with
\begin{equation}
{\cal D}(\omega)=(\omega-\tilde{\epsilon}_{1})(\omega-\tilde{\epsilon}_{2})\pm
i\tilde{\Gamma}_{1}(\omega-\tilde{\epsilon}_{1})\pm i\tilde{\Gamma}_{2}
(\omega-\tilde{\epsilon}_{2}),
\end{equation}
$\tilde{\epsilon}_{j}\equiv\epsilon_{j}+\lambda_{j}$, and
$\tilde{\Gamma}_{j}=\Gamma |z_{j}|^2$, $\Gamma = \pi \sum_{k} |V|^2 \delta(\omega
 -\epsilon_{\eta k})$ being the coupling constant between the QD and the lead.
$f_{\eta}(\omega)$ is the Fermi distribution function of $\eta$th lead ($\eta=L/R$).
The current $I$ through the two-level QD can be divided into three parts: the 
contributions from the ground level $I_{1}$, from the second level $I_{2}$, and the 
current $I_{\rm i}$ resulting from the interference between the two energy levels. 
Straightforwardly, we have $I=I_{1}+I_{2}+I_{\rm i}=\frac{2e}{h}\int_{-\infty}^{+\infty} 
[T_{1}(\omega)+T_{2}(\omega)+T_{\rm i}(\omega)][f_{L}(\omega)-f_{R}(\omega)]$, in which
\bn
T_{1}(\omega)&=&\tilde{\Gamma}_{1}^2 (\omega-\tilde{\epsilon}_{2})^2 |{\cal D} 
(\omega)|^{-2}, \\
T_{2}(\omega)&=&\tilde{\Gamma}_{2}^2 (\omega-\tilde{\epsilon}_{1})^2 |{\cal D} 
(\omega)|^{-2}, \\
T_{\rm i}(\omega)&=&2\tilde{\Gamma}_{1} \tilde{\Gamma}_{2} (\omega-\tilde{\epsilon}_{2}) 
(\omega-\tilde{\epsilon}_{1}) |{\cal D}(\omega)|^{-2}
\en
are the transmission probabilities. 

In the linear limit the conductance can be written as
\bq
G=\frac{2e^2}{h} [T_{1}(0) + T_{2}(0) + T_{\rm i}(0) ] 
=\frac{2e^2}{h}\frac{(\tilde{\Gamma}_{1}\tilde{\epsilon}_{2} + 
\tilde{\Gamma}_{2}\tilde{\epsilon}_{1})^2}{|{\cal D} (0)|^2}.
\eq  
In numerical analysis, we fix the parameters of the QD: $U=4$, $J=2$ with 
$\epsilon_1=-2.5$ or $-3$, to guarantee nearly two electrons dwelling in the QD (for 
$-4<\Delta<1.2$, $1.45\lesssim N\lesssim 1.6$ at $\epsilon_{1}=-2.5$; $1.5\lesssim 
N\lesssim 1.7$ at $\epsilon_{1}=-3$), then focus our attention on the $\Delta$-dependent 
Kondo effect in transport. Here $\Gamma$ is always taken as the energy unit and 
$\epsilon_{j}$ is measured from the Fermi level of leads. Fig.\,1 and the inset figure 
show the total linear conductance $G$ and its three parts as functions of the energy 
splitting $\Delta$ between the spin-singlet and -triplet states in the QD at zero 
temperature. There appears a peak in $G$ near the singlet-triplet degeneracy point 
$\Delta=0$. In the spin-singlet-state-dominated regime (singlet regime), $\Delta<0$, the 
contribution of the second level, $G_{2}$, and that of the interference between the two 
levels, $G_{\rm i}$, are almost zero. Only the first level is the active conduction 
channel. With increasing $\Delta$ (decreasing $\delta$), the second level starts to carry 
current, such that $G$ rises up. Around the degeneracy point, the interference effect 
appears and its contribution $G_{\rm i}$ is negative. The rapid enhancement of the 
interference effect causes an abrupt decline of $G$ in the spin-triplet-state-dominated 
regime (triplet regime) $\Delta>0$ and a sharp peak in $G$ near the transition point. Note 
that the nonzero value of $G$ in the singlet regime is determined by the position in the 
Coulomb blockade valley due to $N<2$ in our model QD parameters.

Fig.\,2 illustrates the behavior of the conductance $G$ as a function of temperature up to 
$0.3$. It is believed that the SBMF approach is correct for describing the Kondo effect in 
strongly correlated systems at temperatures lower than the Kondo temperature $T_{\rm K}$. 
For the systems under consideration, the Kondo temperatures $T_{\rm K}$ are estimated to 
be about $0.21$ and $0.32$ if only the first level is considered. \cite{tk} Therefore we 
can cautiously believe that the present calculation gives a rational description for the 
temperature dependence of the Kondo contribution to the conductance. In the cases 
$\Delta=0$ and $0.5$, the conductance rises monotonously with decreasing temperature. In 
the cases $\Delta=1$ and $1.2$, however, a ``hump" behavior appears. After riseup with 
lowering temperature at first, $G$ reaches a peak at a certain value of $T$ before 
decreases, indicating a two-stage Kondo effect in this triplet regime. These results are 
in qualitatively agreement with the scaling analysis of Pustilnik and Glazman  
\cite{lateral}, where a two-stage Kondo effect for an $S\ge 1$ ground state (i.e., the 
triplet regime in this letter) was suggested. Moreover, we can see from Fig.2 that more 
pronounced hump appears in deeper triplet regime (bigger $\Delta$). It is also worth 
mentioning that, for moderate $\Delta$, for example, $\Delta=1$ at $\epsilon_{1}=-2.5$ 
[Fig.\,2(a)] and $\Delta=0.5$ at $\epsilon_{1}=-3$ [Fig.\,2(b)], the conductance $G$ may 
slightly increase at very low temperatures, exhibiting a ``shoulder" behavior, which 
satisfies very well with the recent measurements in Ref.\,\cite{Wiel} (inset of Fig.\,2).

To simplify the calculation of nonlinear transport under a finite bias voltage between the 
two leads, we assume a symmetric voltage drop, $\mu_L=-\mu_R=eV/2$. The calculated 
zero-temperature current $I$ as a function of bias voltage $V$ in the case of $U=4$, $J=2$ 
and $\epsilon_{1}=-2.5$ are shown in Fig.\,3(a) for $\Delta\geq 0.5$. With increasing the 
bias voltage in the range shown in the figure, 1) the current $I$ increases monotonically 
for $\Delta \lesssim 0.5$; 2) $I$ approaches a peak value followed by a drop for $\Delta > 
0.5$; 3) the peak current appears at smaller bias for bigger $\Delta$. Correspondingly the 
voltage-dependent differential conductance $dI/dV$ [Figs.\,3(b)] shows a 
non-zero-bias-maximum at triplet regime $\Delta > 0.5$. While with decreasing $\Delta$, 
this sharp dip structure gradually transits to a conventional out-of-equilibrium Kondo 
behavior, i.e., zero-bias-maximum, at $\Delta=0.5$. As shown in the insets in Figs.\,2 and 
3(b), where the separate contributions of the first and second levels as well as the 
interference term, to $G$ and $dI/dV$ are shown respectively, the hump behavior and the 
shoulder structure in $G$-$T$ curves, and the sharp dip in $dI/dV$-$V$ curves, can all be 
attributed to the strongly weakening of the interference effect due to increasing 
temperature or bias voltage.           

In conclusion, we have generalized the finite-$U$ SBMF approach to the case of multi-level 
QD, and employed it to investigate the singlet-triplet Kondo phenomena in linear and 
nonlinear transport through a two levels QD in the $N\simeq 2$ regime. Our investigation 
revealed a two-stage Kondo effect, a non-monotonic temperature-dependent conductance and a 
sharp dip structure in differential conductance, depending on the energy splitting 
$\Delta$ between the spin-singlet and triplet states. These are in agreement with recent 
experimental observation. The predicted behavior of linear conductance is in consistent 
with existing theoretical analysis. An interpretation for this two-stage Kondo behavior in 
$G$-$T$ and $dI/dV$-$V$ has been provided in terms of the interference between the two 
levels.

{\it Acknowledgment}. This work was supported by the National Natural Science Foundation
of China (grant Nos. 60076011 and 90103027), the Special Funds for Major
State Basic Research Project (grant No. 2000683), and the
Shanghai Municipal Commission of Science and Technology.

\begin{center}
{\bf Figure Captions}
\end{center}

\vspace{0.2cm}
\noindent
{\bf Fig.1} Linear conductance at the absolute zero temperature as a function of the 
splitting between the spin-singlet and triplet states $\Delta$ in QD of $U=4$ and $J=2$ 
with $\epsilon_{1}=-2.5$ and $-3$. Inset: Corresponding contributions of three parts 
$G_{1}$, $G_{2}$ and $G_{\rm i}$ for $\epsilon_{1}=-3$.

\vspace{0.2cm}
\noindent
{\bf Fig.2} Temperature dependence of linear conductance with $\Delta=1.2$, $1$, $0.5$, 
and $0$ for (a) $\epsilon_{1}=-2.5$ and  (b) $\epsilon_{1}=-3$. Insets: Corresponding 
contributions of three parts $G_{1}$, $G_{2}$ and $G_{\rm i}$ for (a) $\Delta=1$ and for 
(b) $\Delta=1.2$.  

\vspace{0.2cm}
\noindent
{\bf Fig.3} (a) Zero-temperature $I$-$V$ characteristic curves for QDs with $U=4$, $J=2$, 
and $\epsilon_{1}=-2.5$ at $\Delta=1.2\sim 0.5$. (b) Zero-temperature differential 
conductance $dI/dV$ as a function of the external voltage at $\Delta=0.5$, $1.0$, and 
$1.2$. Inset in (b): Corresponding contributions of three parts $dI_{1}/dV$, $dI_{2}/dV$ 
and $dI_{\rm i}/dV$.


\begin{thebibliography}{10}
\bibitem{tk}{The Kondo temperature for the single-level QD is given by $T_{\rm K} 
=U\sqrt{\beta}\exp(-\pi/\beta)/2\pi$, $\beta=-2U\Gamma/\epsilon_{1}(U+\epsilon_{1})$.}
\end{thebibliography}
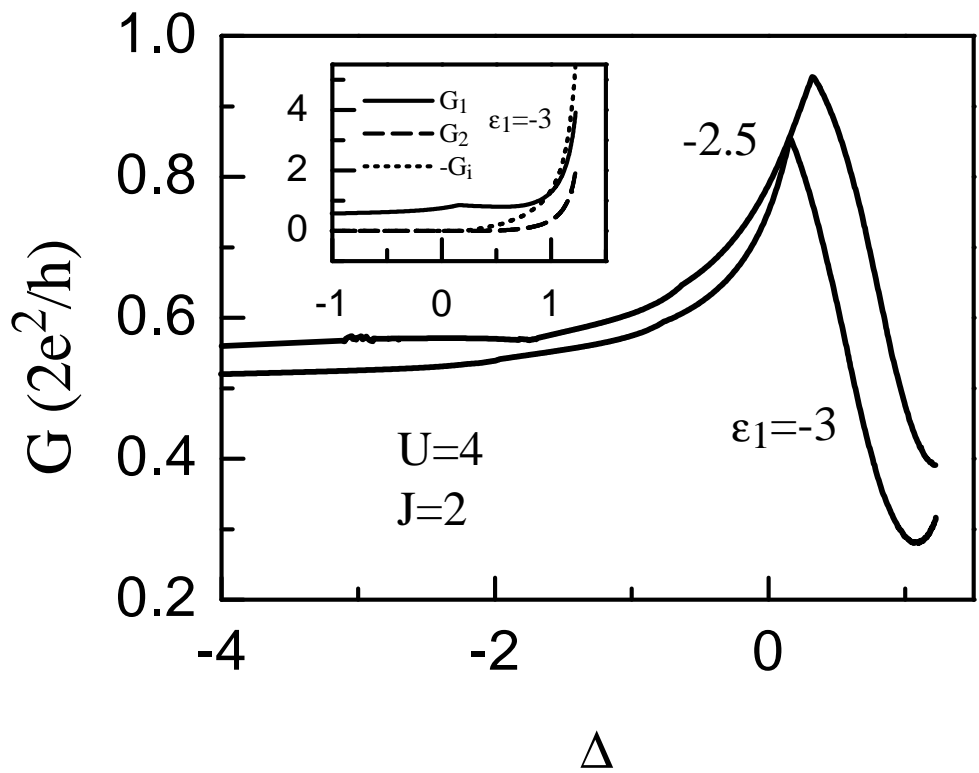
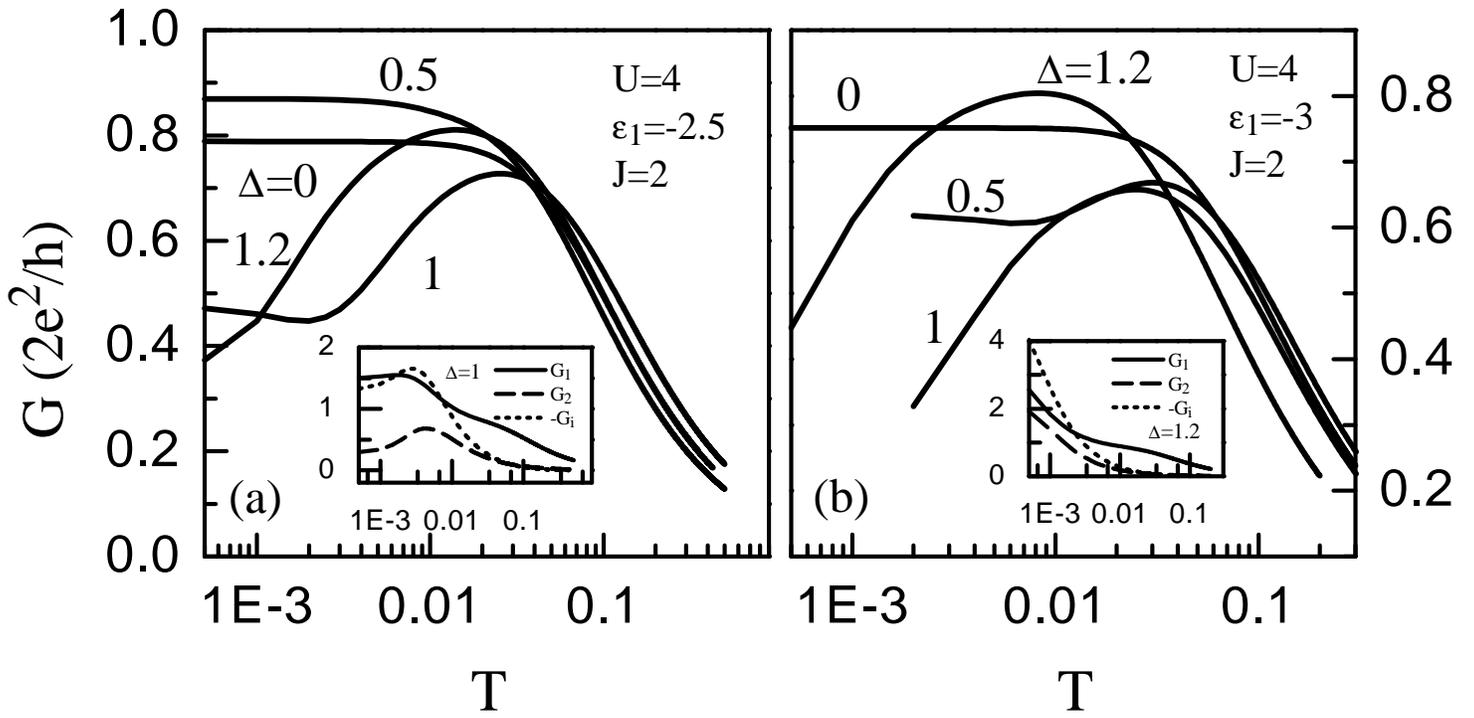
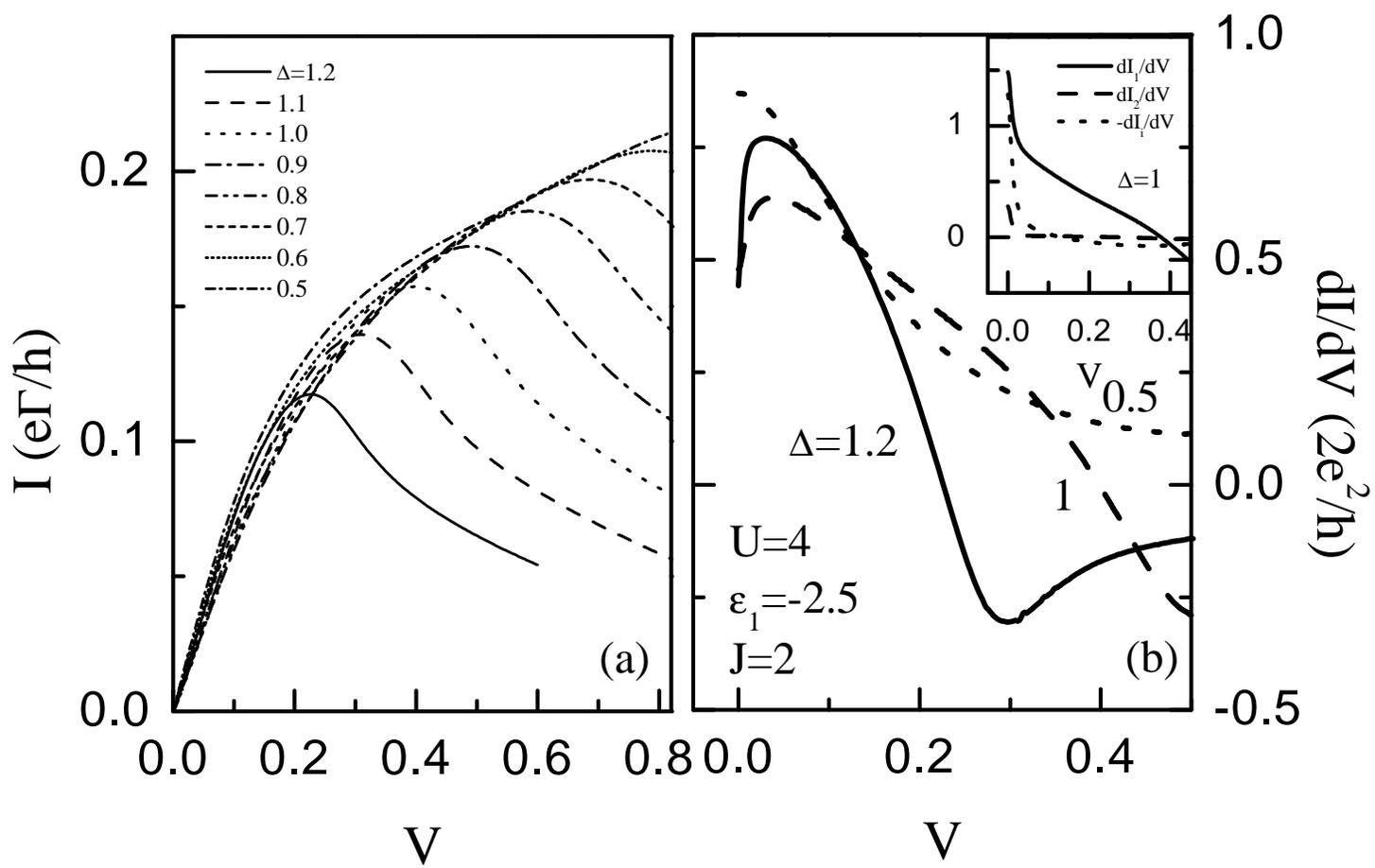
\end{document}